\pdfoutput=1
\documentclass[11pt]{article}

\usepackage[final]{acl}

\usepackage{times}
\usepackage{latexsym}
\usepackage{multirow}
\usepackage{tabularx}

\usepackage[T1]{fontenc}
\usepackage[utf8]{inputenc}

\usepackage{microtype}

\usepackage{inconsolata}

\usepackage{graphicx}

\title{Civil Society in the Loop: Feedback-Driven Adaptation of (L)LM-Assisted Classification in an Open-Source Telegram Monitoring Tool}
\author{
  \textbf{Milena Pustet\textsuperscript{1}},
  \textbf{Elisabeth Steffen\textsuperscript{1}},
  \textbf{Helena Mihaljević\textsuperscript{1}},
  \textbf{Grischa Stanjek\textsuperscript{2}},
   \textbf{Yannies Illies\textsuperscript{2}}
\\
  \textsuperscript{1}HTW Berlin, 
  \textsuperscript{2}democ
  \\\texttt{mihalje@htw-berlin.de}
}

\begin{document}
\maketitle

\section{Introduction}
% deadline: 18.04. AOE
The role of civil society organizations (CSOs) in monitoring harmful online content is increasingly crucial, especially as platform providers reduce their investment in content moderation \cite{kayyali_metas_2025}. 
AI tools can assist in detecting and monitoring harmful content at scale. However, few open-source tools offer seamless integration of AI models and social media monitoring infrastructures. 

Given their thematic expertise and contextual understanding of harmful content, CSOs should be active partners in co-developing technological tools, providing feedback, helping to improve models, and ensuring alignment with  stakeholder needs and values, rather than as passive `consumers'.
However, collaborations between the open source community, academia, and civil society remain rare, and research on harmful content seldom translates into practical tools usable by civil society actors.

This work in progress explores how CSOs can be meaningfully involved in an AI-assisted open-source monitoring tool of anti-democratic movements on Telegram, which we are currently developing in collaboration with CSO stakeholders. 
Telegram has become a key platform for far-right and other extremist groups, enabling fast, large-scale dissemination of harmful, increasingly multi-modal content \cite{baumgartner_pushshift_2020,urman_what_2022}.
Using conspiracy theory classification as a scenario, we discuss two concrete approaches to such collaboration.

\section{Related Work}
Several CSOs in German-speaking countries, such as CeMAS and ISD, use in-house and closed-source monitoring tools for Telegram and other platforms. 
Open-source scripts for Telegram analysis exist, but remain inaccessible to most non-technical users. GESIS offers a web data collection service with a clear focus on academic researchers.

An early contribution by \citet{schinas_open-source_2017} offered a modular, locally deployable monitoring system with basic NLP and relevance feedback, though it is no longer maintained. 
Existing open-source tools such as Telegram Monitor \cite{telegram_monitor_2022} or those developed by \citet{primig_introducing_2024,silva_telegramscrap_2024,ruscica_telecatch_2025} mainly focus on data extraction, search and statistical analyses, while 4CAT \cite{peeters_4cat_2022} integrates NLP techniques and prioritizes transparency. AI-based analysis and user feedback integration are still missing across tools.

In adjacent domains like fact-checking, there is a growing demand for AI assistance, particularly for monitoring social media \cite{juneja_human_2022,hrckova_autonomation_2024}. Despite the availability of capable models (e.\,g., for transcription), their integration remains difficult for non-technical users.
Importantly, fact-checkers and researchers emphasize that AI must complement, not replace human expertise  \cite{procter_observations_2023}, calling for adaptive systems that incorporate user feedback \cite{juneja_human_2022} and ensure data ownership and privacy \cite{wolfe_impact_2024}.

\section{Feedback-Driven Model Integration}

\begin{figure*}[ht!]
    \centering
    \includegraphics[width=0.9\linewidth]{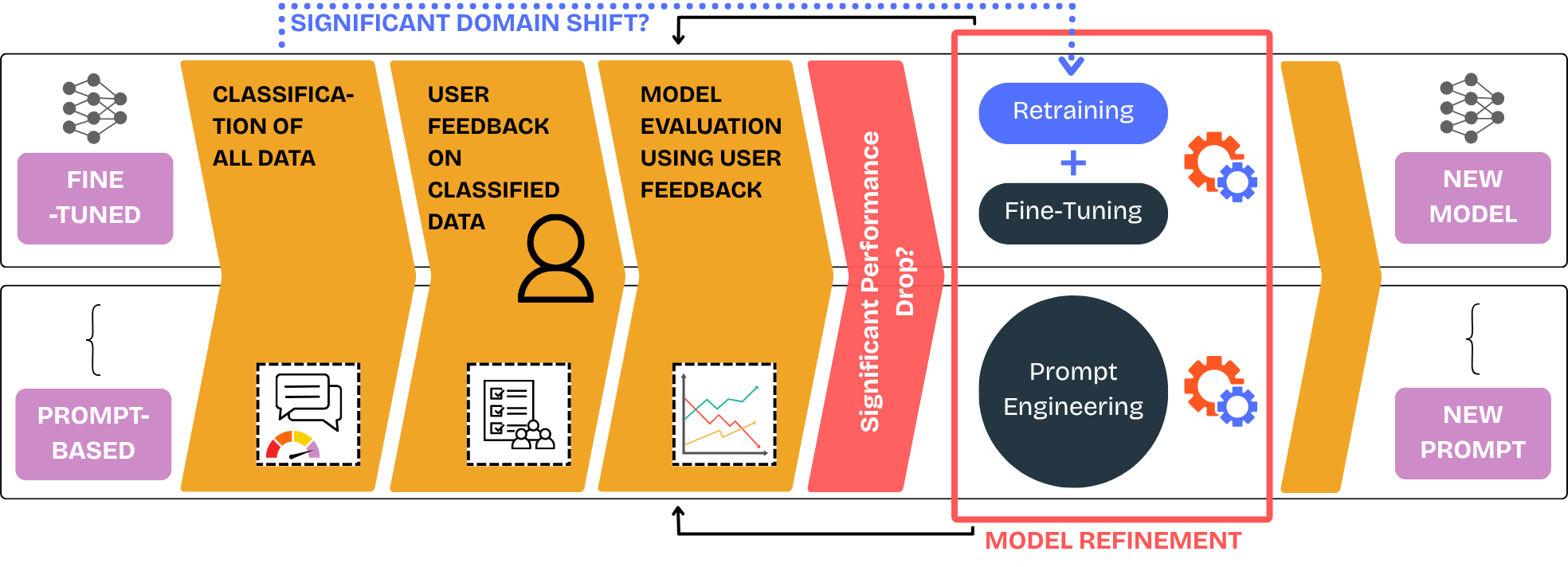}
    \caption{User feedback integration workflow for fine-tuned and prompt-based content classification.}
    \label{fig:workflow}
\end{figure*}

The software is designed for use by individual CSOs in their own dedicated environments. It enables users to search for content within Telegram channels and presents the results in a feed format, where each post is assigned a binary CT classification label along with a corresponding confidence score. Since the classifier's output is only one of several features, its evaluation is not the primary objective but is intended to be embedded within the main monitoring workflow. In line with this design, user feedback should focus on the binary labels and be integrated directly into the search interface.

There are currently two main approaches for integrating classification models into monitoring systems: fine-tuning (FT) a domain-specific model or prompting (P) an LLM. We use a well-performing BERT-based model for CT classification trained on Telegram data \cite{pustet-etal-2024-detection} as a solid starting point for FT-based classification.

Figure~\ref{fig:workflow} outlines a workflow integrating both approaches. During monitoring, users provide feedback on the class labels displayed alongside each message. This feedback, stored with the original classifications and feedback-related metadata, forms a growing gold-standard dataset for evaluation and model refinement.
User feedback supports refinement in both approaches: it informs fine-tuning in FT and guides prompt revision or few-shot updates in P. In the FT pathway, concept drift, detected by comparing new data to the training corpus, can additionally trigger model retraining. Updated models or prompts are then re-evaluated with accumulated feedback and iteratively deployed.

Several challenges must be addressed. First, due to negativity bias, users of the software are more likely to provide feedback when they disagree with a label, while \textbf{agreement often goes unmarked}. However, positive signals are also important for model evaluation and refinement, especially in the FT approach. One option is to collect representative explicit feedback through dedicated data views, in which users are asked to rate a full sample of posts. However, this introduces additional workload and deviates from users' regular monitoring routines.  
An alternative is to infer implicit affirmation by treating posts that were seen or clicked but left unmarked as signals of agreement, potentially weighting different user actions (e.g., scrolling vs. clicking) accordingly. Implementing this strategy would require not only experimental validation but also continuous tracking of which users have seen which posts, posing both technical and data privacy challenges.

Second, when multiple users are involved, \textbf{conflicting feedback must be resolved}. While explicit input should take precedence over implicit signals, disagreements between users may need to be routed to a dedicated conflict resolution interface.

Retraining schedules for FT and prompt revision cycles for P also need to be defined, along with protocols for prompt governance. Both approaches require \textbf{validation before deployment}, using part of the growing gold-standard dataset for test and validation sets. In FT, staged rollouts or A/B testing may be needed; in P, prompt changes can be  applied immediately, but must still be monitored. 

Finally, all components must operate in \textbf{secure, cost-efficient environments}. This requirement limits the use of external services, as sending data to third-party APIs raises privacy and compliance concerns. At the same time, running internal (L)LMs on local infrastructure typically requires GPU-intensive hardware, which most, especially smaller, CSOs do not possess. These constraints suggest that, regardless of the chosen approach, smaller or distilled models should be prioritized to balance performance, cost, and accessibility.

Table~\ref{tab:comparison_finetuning_prompting} compares FT and P across flexibility, control, resources, and adaptability. FT offers stable performance, high output consistency, and control, but retraining is technically demanding, especially in low-resource settings. On the other hand, practical issues such as output parsing and prompt formatting remain challenging in the P approach, particularly in multilingual or noisy settings like Telegram. P is more adaptable to evolving discourse and allows easier task switching. While growing user familiarity with LLM prompting may lower entry barriers, evaluating prompt effectiveness and refining outputs still requires structured feedback, technical guidance, and robust evaluation methods. For example, few-shot prompting highly depends on the number and choice of examples \cite{chae_large_2025,olney_impact_2024}, requiring automated experiments evaluated on the validation set.

\begin{table}[ht]
\centering
\small
\caption{Comparison of FT and P. Symbols: \textbf{++} = very good, \textbf{+} = acceptable, \textbf{0} = limited}
\label{tab:comparison_finetuning_prompting}
\begin{tabularx}{\columnwidth}{X|p{0.35cm}p{0.35cm}}
\hline
 \textbf{Aspect} & \textbf{FT} & \textbf{P} \\
\hline
\multicolumn{3}{c}{\textbf{Flexibility}} \\\hline
    Adaptability to task variations  & 0  & ++ \\
    Handling long input                 & 0  & ++ \\
    Flexibility across task scopes     & 0  & ++ \\
    Robustness to raw inputs    & 0  & ++ \\
\hline
\multicolumn{3}{c}{\textbf{Control \& Output}}\\\hline
    Output consistency           & ++ & 0 \\
    Explainability               & 0  & + \\
    Model behavior control (incl. reproducibility)       & ++ & 0 \\
\hline
\multicolumn{3}{c}{\textbf{Resources}} \\\hline
    Model size                   & ++ & 0/+ \\
    Hardware requirements        & ++ & 0/+ \\
    (Re-)training and fine-tuning & 0  & ++ \\
    Human effort required  & 0  & + \\    
\hline
\multicolumn{3}{c}{\textbf{Adaptation \& Evaluation}}\\\hline
    Adaptability through user feedback & +  & 0 \\
        Ease of evaluation           & + & + \\
\hline
\end{tabularx}
\end{table}

\section{Future Work}

We are currently prototyping the integration of both FT and P approaches into the Telegram monitoring software, addressing the concerns outlined above. Ongoing stakeholder research will help align the final implementation with the practical needs and constraints of CSO partners.

Future work should explore hybrid strategies such as pre-labeling messages with high-confidence outputs while directing uncertain cases to users for validation. Additional directions include support for prompt optimization and extending prompting to tasks beyond classification, such as summarization or question answering, though these require substantial user feedback and careful evaluation design. Research into sharing annotated data or federating models across CSOs would also be valuable in the long term. While such strategies may improve scalability, they necessitate robust governance mechanisms to manage potential divergences in labeling practices and organizational priorities.

\bibliography{custom}

\begin{thebibliography}{16}
\providecommand{\natexlab}[1]{#1}

\bibitem[{Baumgartner et~al.(2020)Baumgartner, Zannettou, Squire, and Blackburn}]{baumgartner_pushshift_2020}
Jason Baumgartner, Savvas Zannettou, Megan Squire, and Jeremy Blackburn. 2020.
\newblock \href {https://doi.org/10.1609/icwsm.v14i1.7348} {The {Pushshift} {Telegram} {Dataset}}.
\newblock \emph{Proceedings of the International AAAI Conference on Web and Social Media}, 14:840--847.

\bibitem[{Chae and Davidson(2025)}]{chae_large_2025}
Youngjin Chae and Thomas Davidson. 2025.
\newblock \href {https://doi.org/10.1177/00491241251325243} {Large {Language} {Models} for {Text} {Classification}: {From} {Zero}-{Shot} {Learning} to {Instruction}-{Tuning}}.
\newblock \emph{Sociological Methods \& Research}, page 00491241251325243.

\bibitem[{Hrckova et~al.(2024)Hrckova, Moro, Srba, Simko, and Bielikova}]{hrckova_autonomation_2024}
Andrea Hrckova, Robert Moro, Ivan Srba, Jakub Simko, and Maria Bielikova. 2024.
\newblock \href {https://doi.org/10.48550/arXiv.2211.12143} {Autonomation, not {Automation}: {Activities} and {Needs} of {Fact}-checkers as a {Basis} for {Designing} {Human}-{Centered} {AI} {Systems}}.
\newblock \emph{arXiv preprint}.
\newblock ArXiv:2211.12143.

\bibitem[{Juneja and Mitra(2022)}]{juneja_human_2022}
Prerna Juneja and Tanushree Mitra. 2022.
\newblock \href {https://doi.org/10.1145/3555143} {Human and {Technological} {Infrastructures} of {Fact}-checking}.
\newblock \emph{Proc. ACM Hum.-Comput. Interact.}, 6(CSCW2):418:1--418:36.

\bibitem[{J\'{u}nior et~al.(2022)J\'{u}nior, Melo, Kansaon, Mafra, Sa, and Benevenuto}]{telegram_monitor_2022}
Manoel J\'{u}nior, Philipe Melo, Daniel Kansaon, Vitor Mafra, Kaio Sa, and Fabricio Benevenuto. 2022.
\newblock \href {https://doi.org/10.1145/3511095.3536375} {Telegram monitor: Monitoring brazilian political groups and channels on telegram}.
\newblock In \emph{Proceedings of the 33rd ACM Conference on Hypertext and Social Media}, HT '22, page 228–231, New York, NY, USA. Association for Computing Machinery.

\bibitem[{Kayyali(2025)}]{kayyali_metas_2025}
Dia Kayyali. 2025.
\newblock \href {https://techpolicy.press/metas-content-moderation-changes-are-going-to-have-a-real-world-impact-its-not-going-to-be-good} {Meta's {Content} {Moderation} {Changes} are {Going} to {Have} a {Real} {World} {Impact}. {It}'s {Not} {Going} to be {Good}. {\textbar} {TechPolicy}.{Press}}.

\bibitem[{Peeters and Hagen(2022)}]{peeters_4cat_2022}
Stijn Peeters and Sal Hagen. 2022.
\newblock \href {https://doi.org/10.5117/CCR2022.2.007.HAGE} {The {4CAT} {Capture} and {Analysis} {Toolkit}: {A} {Modular} {Tool} for {Transparent} and {Traceable} {Social} {Media} {Research}}.
\newblock \emph{Computational Communication Research}, 4(2):571--589.

\bibitem[{Primig and Fröschl(2024)}]{primig_introducing_2024}
Florian Primig and Fabian Fröschl. 2024.
\newblock \href {https://doi.org/10.1177/20501579241244973} {Introducing the {FROG} tool for gathering {Telegram} data}.
\newblock \emph{Mobile Media \& Communication}, 12(2):449--453.

\bibitem[{Procter et~al.(2023)Procter, Arana-Catania, He, Liakata, Zubiaga, Kochkina, and Zhao}]{procter_observations_2023}
Rob Procter, Miguel Arana-Catania, Yulan He, Maria Liakata, Arkaitz Zubiaga, Elena Kochkina, and Runcong Zhao. 2023.
\newblock \href {https://doi.org/10.48550/arXiv.2305.02224} {Some {Observations} on {Fact}-{Checking} {Work} with {Implications} for {Computational} {Support}}.
\newblock \emph{arXiv preprint}.
\newblock ArXiv:2305.02224.

\bibitem[{Pustet et~al.(2024)Pustet, Steffen, and Mihaljevic}]{pustet-etal-2024-detection}
Milena Pustet, Elisabeth Steffen, and Helena Mihaljevic. 2024.
\newblock \href {https://doi.org/10.18653/v1/2024.woah-1.2} {Detection of conspiracy theories beyond keyword bias in {G}erman-language telegram using large language models}.
\newblock In \emph{Proceedings of the 8th Workshop on Online Abuse and Harms (WOAH 2024)}, pages 13--27, Mexico City, Mexico. Association for Computational Linguistics.

\bibitem[{Ruscica et~al.(2025)Ruscica, Tucci, and Carneiro}]{ruscica_telecatch_2025}
Giosuè Ruscica, Giulia Tucci, and Bia Carneiro. 2025.
\newblock \href {https://doi.org/10.1016/j.simpa.2024.100736} {{TeleCatch}: {An} open-access software for visualizing, filtering and extracting {Telegram} messages data}.
\newblock \emph{Software Impacts}, 23:100736.

\bibitem[{Schinas et~al.(2017)Schinas, Papadopoulos, Apostolidis, Kompatsiaris, and Mitkas}]{schinas_open-source_2017}
Manos Schinas, Symeon Papadopoulos, Lazaros Apostolidis, Yiannis Kompatsiaris, and Pericles~A. Mitkas. 2017.
\newblock \href {https://doi.org/10.1007/978-3-319-70284-1_28} {Open-{Source} {Monitoring}, {Search} and {Analytics} {Over} {Social} {Media}}.
\newblock In \emph{Internet {Science}}, pages 361--369, Cham. Springer International Publishing.

\bibitem[{Silva(2024)}]{silva_telegramscrap_2024}
Ergon Cugler de~Moraes Silva. 2024.
\newblock \href {https://doi.org/10.48550/arXiv.2412.16786} {{TelegramScrap}: {A} comprehensive tool for scraping {Telegram} data}.
\newblock \emph{arXiv preprint}.
\newblock ArXiv:2412.16786.

\bibitem[{Urman and Katz(2022)}]{urman_what_2022}
Aleksandra Urman and Stefan Katz. 2022.
\newblock \href {https://doi.org/10.1080/1369118X.2020.1803946} {What they do in the shadows: examining the far-right networks on {Telegram}}.
\newblock \emph{Information, Communication \& Society}, 25(7):904--923.

\bibitem[{Wolfe and Mitra(2024)}]{wolfe_impact_2024}
Robert Wolfe and Tanushree Mitra. 2024.
\newblock \href {https://doi.org/10.1145/3630106.3658987} {The {Impact} and {Opportunities} of {Generative} {AI} in {Fact}-{Checking}}.
\newblock In \emph{The 2024 {ACM} {Conference} on {Fairness}, {Accountability}, and {Transparency}}, pages 1531--1543.

\bibitem[{Yoshida(2024)}]{olney_impact_2024}
Lui Yoshida. 2024.
\newblock \href {https://doi.org/10.1007/978-3-031-64315-6_5} {The {Impact} of {Example} {Selection} in {Few}-{Shot} {Prompting} on {Automated} {Essay} {Scoring} {Using} {GPT} {Models}}.
\newblock In Andrew~M. Olney, Irene-Angelica Chounta, Zitao Liu, Olga~C. Santos, and Ig~Ibert Bittencourt, editors, \emph{Artificial {Intelligence} in {Education}. {Posters} and {Late} {Breaking} {Results}, {Workshops} and {Tutorials}, {Industry} and {Innovation} {Tracks}, {Practitioners}, {Doctoral} {Consortium} and {Blue} {Sky}}, volume 2150, pages 61--73. Springer Nature Switzerland, Cham.

\end{thebibliography}

\end{document}